%% file: sample-sigconf.tex
\documentclass[sigconf]{acmart}
\AtBeginDocument{%
  }

\copyrightyear{2025}
\acmYear{2025}
\setcopyright{acmlicensed}\acmConference[KDD '25]{Proceedings of the 31st ACM SIGKDD Conference on Knowledge Discovery and Data Mining V.2}{August 3--7, 2025}{Toronto, ON, Canada}
\acmBooktitle{Proceedings of the 31st ACM SIGKDD Conference on Knowledge Discovery and Data Mining V.2 (KDD '25), August 3--7, 2025, Toronto, ON, Canada}
\acmDOI{10.1145/3711896.3737192}
\acmISBN{979-8-4007-1454-2/2025/08}

\usepackage{hyperref}
\usepackage{booktabs}
\usepackage{multirow}
\usepackage{algorithm}
\usepackage{algorithmic}
\usepackage{enumitem}
\usepackage{enumerate}
\usepackage{xcolor}
\usepackage{graphicx}
\usepackage{subcaption}
\usepackage{diagbox} 
\usepackage{makecell}
\usepackage{array}

\newcommand{\modelname}{AutoGraph}

\newcommand{\eg}{\emph{e.g.}}
\newcommand{\ie}{\emph{i.e.}}

\begin{document}

\title{An Automatic Graph Construction Framework based on Large Language Models for Recommendation}

\author{Rong Shan}
\email{shanrong@sjtu.edu.cn}
\affiliation{
  \institution{Shanghai Jiao Tong University}
  \city{Shanghai}
  \country{China}
}

\author{Jianghao Lin}
\email{chiangel@sjtu.edu.cn}
\authornote{Jianghao Lin is the corresponding author.}
\affiliation{
  \institution{Shanghai Jiao Tong University}
  \city{Shanghai}
  \country{China}
}

\author{Chenxu Zhu}
\email{zhuchenxu1@huawei.com}
\affiliation{
  \institution{Huawei Noah's Ark Lab}
  \city{Shanghai}
  \country{China}
}

\author{Bo Chen}
\email{chenbo116@huawei.com}
\affiliation{
  \institution{Huawei Noah's Ark Lab}
  \city{Shanghai}
  \country{China}
}

\author{Menghui Zhu}
\email{zhumenghui1@huawei.com}
\affiliation{
  \institution{Huawei Noah's Ark Lab}
  \city{Shanghai}
  \country{China}
}

\author{Kangning Zhang}
\email{zhangkangning@sjtu.edu.cn}
\affiliation{
  \institution{Shanghai Jiao Tong University}
  \city{Shanghai}
  \country{China}
}

\author{Jieming Zhu}
\email{jiemingzhu@ieee.org}
\affiliation{
  \institution{Huawei Noah's Ark Lab}
  \city{Shenzhen}
  \country{China}
}

\author{Ruiming Tang}
\email{tangruiming@huawei.com}
\affiliation{
  \institution{Huawei Noah's Ark Lab}
  \city{Shenzhen}
  \country{China}
}

\author{Yong Yu}
\email{yyu@apex.sjtu.edu.cn}
\affiliation{
  \institution{Shanghai Jiao Tong University}
  \city{Shanghai}
  \country{China}
}

\author{Weinan Zhang}
\email{wnzhang@sjtu.edu.cn}
\affiliation{
  \institution{Shanghai Jiao Tong University}
  \city{Shanghai}
  \country{China}
}

\renewcommand{\shortauthors}{Rong Shan et al.}

\begin{abstract}
Graph neural networks (GNNs) have emerged as state-of-the-art methods to learn from graph-structured data for recommendation. 
However, most existing GNN-based recommendation methods focus on the optimization of model structures and learning strategies based on pre-defined graphs, neglecting the importance of the \textit{graph construction} stage. 
Earlier works for graph construction usually rely on specific rules or crowdsourcing, which are either too simplistic or too labor-intensive. 
Recent works start to utilize large language models (LLMs) to automate the graph construction, in view of their abundant open-world knowledge and remarkable reasoning capabilities. 
Nevertheless, they generally suffer from two limitations: (1) \textit{invisibility of global view} (\eg, overlooking contextual information) and (2) \textit{construction inefficiency}. 
To this end, we introduce \textbf{\modelname}, an automatic graph construction framework based on LLMs for recommendation. 
Specifically, we first use LLMs to infer the user preference and item knowledge, which is encoded as semantic vectors. 
Next, we employ vector quantization to extract the latent factors from the semantic vectors.
The latent factors are then incorporated as extra nodes to link the user/item nodes, resulting in a graph with in-depth global-view semantics.
We further design metapath-based message aggregation to effectively aggregate the semantic and collaborative information. 
The framework is model-agnostic and compatible with different backbone models. 
Extensive experiments on three real-world datasets demonstrate the efficacy and efficiency of \modelname\ compared to existing baseline methods. 
We have deployed \modelname\ in Huawei advertising
platform, and gain a 2.69\% improvement on RPM and a 7.31\% improvement on eCPM in the online A/B test. Currently \modelname\ has been used as the main
traffic model, serving hundreds of millions of people.


\end{abstract}



\maketitle

\input{text/new_intro}

\input{text/formulation}

\input{text/new_method}
\input{text/experiment}

\input{text/related_work}

\input{text/conclusion}

\begin{acks}
The Shanghai Jiao Tong University team is partially supported by National Key R\&D Program of China (2022ZD0114804), Shanghai Municipal Science and Technology Major Project (2021SHZDZX0102) and National Natural Science Foundation of China (624B2096, 62322603, 62177033).
The work is sponsored by Huawei Innovation Research Program.
We thank MindSpore~\cite{mindspore} for the partial support of this work, which is a new deep learning computing framework.
\end{acks}

\bibliographystyle{ACM-Reference-Format}
\bibliography{acmart}

\input{text/appendix}


\end{document}

%% file: text/new_intro.tex
\section{Introduction}
\label{sec:intro}

Recommender systems (RSs) have become increasingly indispensable to alleviate the information overload problem~\cite{dai2021adversarial,fu2023f} and match users’ information needs~\cite{guo2017deepfm,lin2023map,song2012survey} for various online services~\cite{goyani2020review,schafer2001commerce}. 
In the past decades, researchers have proposed various advanced deep learning methodologies to incorporate various model structures~\cite{liu2024mamba4rec,zhou2018deep,lin2021graph} and auxiliary information~\cite{xi2023bird,li2024survey,zhou2023comprehensive} for recommendation. 
Among them, graph neural network (GNN) based methods turn out to be the state-of-the-art algorithms in mining the complex topological distributions from graph-structured relational data for recommender systems~\cite{de2024personalized, gao2023survey, xv2023commerce}.

However, most of the existing GNN-based recommendation methods primarily focus on optimizing the model structures~\cite{shi2018heterogeneous,wang2023knowledge} and learning strategies~\cite{jiang2023adaptive,wu2021self} based on the pre-defined graphs, generally neglecting the importance of the graph construction stage. 
The quality of the graph structure is the foundation for the entire graph learning process, and can directly influence the model's ability to capture the underlying relationships and patterns within the data~\cite{qiao2018data,zhong2023comprehensive}. 
Earlier works~\cite{liu2023understanding, xv2023commerce, } for graph construction usually employ specific rules (\eg, click as linkage, or entity extraction) or human efforts via crowdsourcing (\eg, relational annotation for knowledge graphs).
They are either too simplistic to model the sophisticated semantic signals in the recommendation data, or too labor-intensive to be scalable for large-scale scenarios. 

Nowadays, large language models (LLMs) emerge as a promising solution for automatic graph construction in view of their vast amount of open-world knowledge, as well as their remarkable language understanding and reasoning capabilities~\cite{yang2024give}. 
Recent attempts have proposed various innovative prompt-based techniques, \eg, chain-of-thought prompting~\cite{zhao2024breaking}, multi-turn conversation~\cite{wei2023zero}, and proxy code generation~\cite{bi2024codekgc}, to estimate the relational linkage between nodes for graph construction. 
Although these works automate the graph construction stage with the help of LLMs and largely save the intensive human labor, they still suffer from the following two limitations, especially when faced with the large-scale user/item volumes in industrial recommender systems.

\begin{figure}[t]
  \centering
  \includegraphics[width=0.44\textwidth]{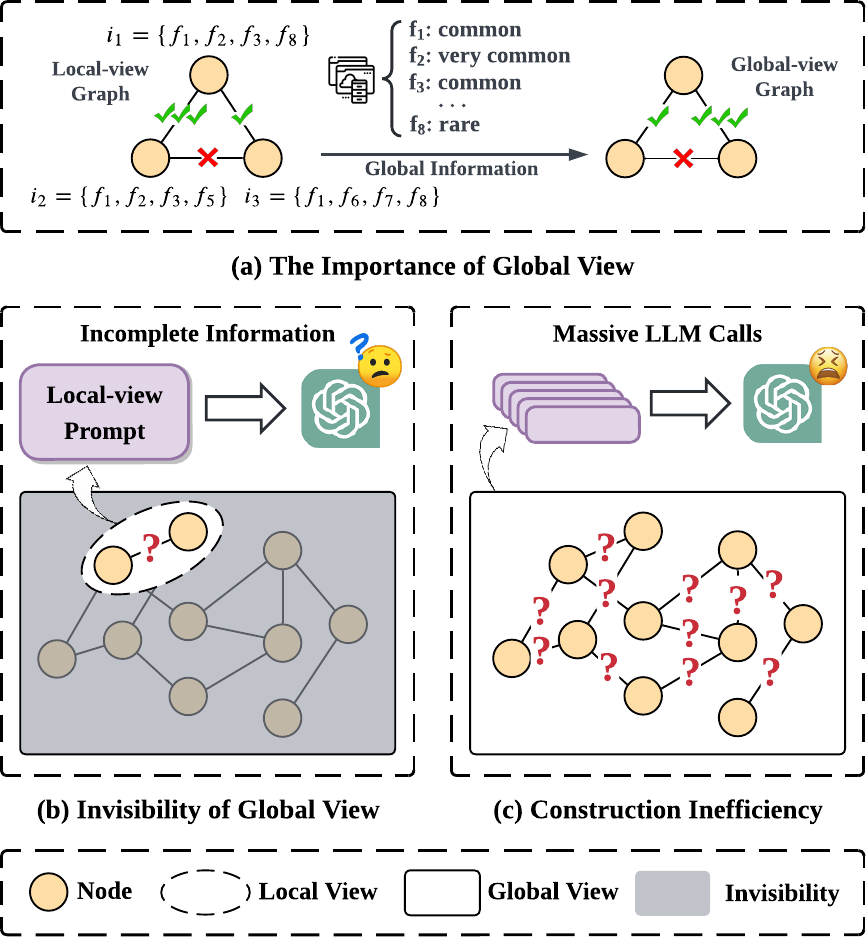}
  \vspace{-5pt}
  \caption{The illustration of (a) the importance of global-view information (\eg, global feature frequency) that can change the optimal graph structure, and the two limitations of existing LLM-based graph construction methods including (b) invisibility of global view, and (c) construction inefficiency.
  }
  \vspace{-10pt}
  \label{fig:illustration}
\end{figure}

\textit{Firstly, existing LLM-based graph construction methods fail to capture the high-quality topological structure among nodes due to the \textbf{invisibility of global view}}. 
The reasonable assessment of node connections should comprehensively consider the global view of the entire dataset, including but not limited to node attributes, graph schema, and contextual information~\cite{yang2024give,ding2024automated}. 
For example, as depicted in Figure~\ref{fig:illustration}(a), suppose we have three item nodes with features: $i_1=\{f_1,f_2,f_3,f_8\}$, $i_2=\{f_1,f_2,f_3,f_5\}$, and $i_3=\{f_1,f_6,f_7,f_8\}$. 
With a simple local-view pairwise comparison, it seems that item $i_1$ is more similar to item $i_2$ since they have more overlapped features.
But once we acquire the global information that $f_8$ serves as a rare feature with fairly low frequency and others are commonly frequent ones, the association between $i_1$ and $i_3$ will be significantly enhanced and the connection between $i_1$ and $i_2$ could be in turn reduced.
Nevertheless, as shown in Figure~\ref{fig:illustration}(b), due to the context window limitation of LLMs and the large-scale users/items in RSs, it is hard to incorporate all the important information into the prompt, which will be truncated and incomplete. Therefore, the information utilized by these methods can only be partial, but never global. Such local-view information can thereby lead to inferior topological graph structures.

\textit{Secondly, existing works generally suffer from the \textbf{construction inefficiency} issue due to the massive invocations of LLMs}. 
While LLMs provide support for in-depth semantic analysis and complex topological structure mining, their intrinsic expensive inference cost poses a significant challenge to the efficiency of graph construction algorithms. 
Most works instruct LLMs to infer the similarity scores between nodes in a pairwise manner~\cite{shang2024survey}, and result in a time complexity of $O(N^2)$, which is impractical for real-world scenarios where the number of users/items $N$ can easily reach million or even billion level~\cite{lin2023map}. 
Although several works propose to conduct downsampling~\cite{sun2023large, wei2024llmrec} or heuristic pre-filtering~\cite{zhao2024breaking} to reduce the number of calls of LLMs, they generally sacrifice the graph quality and thereby introduce noise to the constructed graph. 
Therefore, it is crucial to design an efficient yet effective LLM-automated graph construction method for large-scale industrial applications.

To this end, we propose \textbf{\modelname}, an automatic graph construction framework based on large language models for recommendation. Specifically, \modelname\ consists of two stages: \textit{quantization-based graph construction} and \textit{graph-enhanced recommendation}. 
In the \textit{quantization-based graph construction} stage, we first leverage LLMs to infer the user preference and item knowledge, which is encoded as semantic vectors. 
Such a pointwise invocation manner (\ie, invoking LLMs for each single user/item separately) improves the efficiency by reducing the calls of LLMs to $O(N)$ complexity.
Then we propose latent factor extraction for users and items based on vector quantization techniques. By incorporating the latent factors as extra nodes, we build a graph with a global view of in-depth semantics, providing more comprehensive and informative insights through the topological structure. In the \textit{graph-enhanced recommendation} stage, we propose metapath-based message propagation to aggregate the semantic and collaborative information effectively on the constructed graph, resulting in the graph-enhanced user/item representations. 
These representations can be integrated into arbitrary recommender systems for enhancement.

The main contributions of this paper are as follows:
\begin{itemize}[leftmargin=10pt]
    \item To the best of our knowledge, we are the first to introduce vector quantization for graph construction based on LLMs in recommendation, which addresses the two key limitations of existing methods, \ie, invisibility of global view and inefficiency.
    \item We propose a novel \modelname\ framework, which achieves both effectiveness and efficiency. We extract the latent factors of LLM-enhanced user/item semantics based on vector quantization, which are involved as extra nodes for global-view graph construction.
    Moreover, metapath-based message propagation is designed to aggregate the semantic and collaborative information for recommendation enhancement.
    \item \modelname\ is a general model-agnostic graph construction framework. It is compatible with various recommendation models, and can be easily plugged-in for existing recommender systems.
    \item Experiments on three public datasets validate the superiority of \modelname\ compared to existing baselines. We deploy \modelname\ on an industrial platform, and gain a 2.69\% improvement on RPM and a 7.31\% improvement on eCPM in online A/B test.
\end{itemize}


%% file: text/formulation.tex
\section{Preliminaries}
\label{sec:preliminary}
Given the user set $\mathcal{U}$ and item set $\mathcal{I}$, each user $u \in \mathcal{U}$ has a chronological interaction sequence $ \mathcal{S}^u = [i_l]_{l=1}^L$, where $i_l \in \mathcal{I}$ is the $l$-th item interacted by the user $u$ and $L$ is the length of the interaction sequence. 
Besides, each user $u$ has a profile of multiple features, such as user ID and age, while each item has multiple attributes, such as item ID and genre. 
We can denote them as $\mathcal{F}^u = \{f^u_j\}_{j=1}^{F^u}$ and $\mathcal{F}^i = \{f^i_j\}_{j=1}^{F^i}$, where $F^u$ and $F^i$ denote the number of features for the user and item respectively. 
A typical recommendation model learns a function $\Phi$ to predict the preference score of a user $u$ towards a target item $i$, which can be formulated as:
\begin{equation}
\begin{aligned}
score &= \Phi (\mathcal{S}^u, \mathcal{F}^u, \mathcal{F}^i)~,
\end{aligned}
\label{eq:ori rec score}
\end{equation}
where the model can be optimized for downstream tasks like click-through-rate estimation~\cite{lin2023map,lin2024rella}, or next item prediction~\cite{liu2024mamba4rec,wang2020next}.

As for graph-enhanced recommendation, we will further construct and incorporate a graph $\mathcal{G}=\{\mathcal{V},\mathcal{E}\}$ into the model:
\begin{equation}
\begin{aligned}
score &= \Phi (\mathcal{S}^u, \mathcal{F}^u, \mathcal{F}^i, \mathcal{G}),
\end{aligned}
\label{eq:graph rec score}
\end{equation}
where $\mathcal{V}=\{\mathcal{U},\mathcal{I}\}$ is the set of user nodes $\mathcal{U}$ and item nodes $\mathcal{I}$, and the edge set $\mathcal{E}$ usually contains three types of edges~\cite{he2020lightgcn, gao2023survey}\footnote{For simplicity, we use $\mathcal{U}$ and $\mathcal{I}$ to represent the universal sets of users and items, as well as their corresponding node sets in graph.}:
\begin{itemize}[leftmargin=10pt]
    \item \textbf{User-Item Edge $e_{u-i}$} denotes that the item $i$ is positively interacted by user $u$, \eg, click or purchase. 
    \item \textbf{User-User Edge $e_{u-u}$} indicates the relationship between each pair of users, \eg, social network.
    \item \textbf{Item-Item Edge $e_{i-i}$} represents the similarity between each pair of items, possibly measured by their attributes and contents.
\end{itemize}
Note that there are many graph variants or special cases based on such a basic formulation. For example, many works simply focus on a specific homogeneous graph (\eg, user social networks~\cite{yuan2022semantic}), or the bipartite user-item interaction graph~\cite{wang2019neural} for recommendation. 
Knowledge graphs further introduce the entity nodes and diverse relation edges for item semantic modeling.
In this work, we extend such a basic graph formulation with newly introduced latent factor nodes for both users and items, which will be discussed in Section~\ref{sec:method}.

%% file: text/new_method.tex
\section{Methodology}
\label{sec:method}

\begin{figure*}[t]
    \vspace{-7pt}
  \centering
  \includegraphics[width=0.95\textwidth]{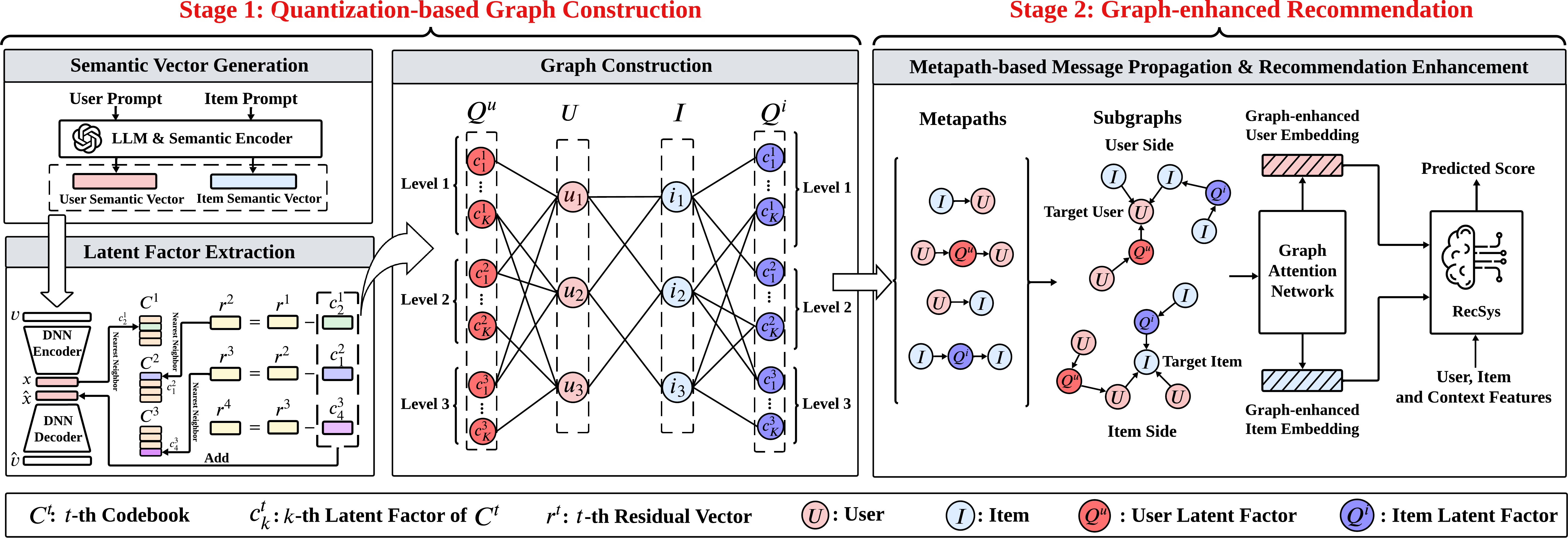}
  \vspace{-8pt}
  \caption{The overall framework of our proposed \modelname.
  }
  \vspace{-10pt}
  \label{fig:framework}
\end{figure*}


\subsection{Overview of \modelname}

As illustrated in Figure~\ref{fig:framework}, our proposed \modelname\ consists of two major stages: (1) quantization-based graph construction, and (2) graph-enhanced recommendation.

\textbf{Quantization-based Graph Construction}. In this stage, we enrich user/item semantics based on LLMs, and leverage vector quantization techniques for a global-view graph construction. With the extracted latent factors as the bridge, we can capture the in-depth semantics and establish the graph with a global learning view, providing more comprehensive and informative insights through the graph structure. 
Besides, by invoking LLMs for each single user/item separately in a pointwise manner, we reduce the calls of LLMs to $O(N)$ complexity and thereby improve the efficiency.

\textbf{Graph-enhanced Recommendation} In this stage, to effectively aggregate the semantic and collaborative information in the constructed graph, we design several metapaths and conduct message propagation based on them. Since our framework is model-agnostic, the obtained graph-enhanced representations can be integrated into arbitrary recommender systems in various way for recommendation enhancement.

\subsection{Quantization-based Graph Construction}

We aim to employ large language models to capture the in-depth semantic signals and complex relationship among users and items for graph construction. 
However, as discussed in Section~\ref{sec:intro}, existing LLM-based graph construction methods generally utilize various prompt techniques to conduct pairwise assessments between each node pair, suffering from the invisibility of global view and construction inefficiency with a time complexity of $O(N^2)$~\cite{sun2023large, chen2024relation}. 

To address these challenges, as shown in Figure~\ref{fig:framework}, we design a quantization-based graph construction method consisting of three steps. 
(1) In the \textit{semantic vector generation} step, we first leverage LLMs to infer user preferences and item knowledge based on their profiles and attributes. 
The LLM-inferred knowledge is then encoded into semantic vectors.
This pointwise invocation manner reduces the calls of LLMs to $O(N)$ complexity and thereby improves the efficiency. 
(2) In the \textit{latent factor extraction} step, we employ vector quantization with latent factors for global-view graph learning, and assign each user/item to a set of factors.  
(3) In the \textit{graph construction} step, we extend the basic user-item graph introduced in Section~\ref{sec:preliminary} by regarding each user/item latent factor as an extra node, resulting in a graph that not only captures in-depth semantics deduced by LLMs, but also retains the global contextual information.

\subsubsection{Semantic Vector Generation}
\label{sec:Semantic Vector Generation}
The semantic information in the recommendation data is often unilateral and shallow~\cite{lin2023can}, and thereby needs to be enriched for graph construction to better capture in-depth semantics.
To this end, we first harness LLMs to infer the user preference and item knowledge based on their vanilla profiles or attributes. We then encode the generated knowledge into semantic vectors $\{v^u_j\}_{j=1}^{|\mathcal{U}|}$ and $\{v^i_j\}_{j=1}^{|\mathcal{I}|}$ for subsequent graph construction. 
Notably, this process is conducted in a pointwise manner (\ie, invoking LLMs for each single user/item separately), and reduces the calls of LLMs to $O(N)$ complexity, which is more efficient compared with the $O(N^2)$ complexity of existing LLM-based graph construction methods~\cite{guo2024graphedit, sun2023large}. 
The process can be formulated as:
\begin{equation}
\begin{aligned}
    v_j^u &= Encoder(LLM(\mathcal{T}_{j}^{u})) \,\in\mathbb{R}^{D_v}, \\
    v_j^i &= Encoder(LLM(\mathcal{T}_{j}^{i})) \,\in\mathbb{R}^{D_v},
\end{aligned}
\end{equation}
where $\mathcal{T}_{j}^{u}$ and $\mathcal{T}_{j}^{i}$ is the prompt for the $j$-th user and item, and $D_v$ is the output dimension. 
The detailed prompt is provided in Appendix~\ref{app: prompt} due to the page limitation. Note that if the LLM is open-source, we can directly adopt the representations from the last hidden layer as the semantic vectors (\ie, LLM as encoder).



\subsubsection{Latent Factor Extraction}
\label{sec: Latent Factor Extraction}
The semantic vectors contain diverse yet noisy information for downstream tasks~\cite{xi2023towards}. 
Moreover, since these semantic vectors are generated in a pointwise manner, it is non-trivial to leverage them for graph construction with a global view, \ie, being aware of the overall connections instead of just the neighborhood linkages. 
To this end, we introduce the latent factors for users and items based on vector quantization techniques. 
These latent factors have multifaceted and interpretable meanings, and showcase the potential connections among users/items, which allows us to employ the global semantic relevance for graph construction. 
Next, we will dive into the details of the quantization techniques for latent factor extraction.
Note that here we omit the superscripts $u$ and $i$ which distinguish users and items for simplicity, as the process is similar for both.

As shown in Figure~\ref{fig:framework}, we employ residual quantization~\cite{rajput2024recommender, liu2024vector} for latent factor extraction. 
We quantize the semantic vectors $v$ using $T$ codebooks denoted as ${\{\mathcal{C}^t\}_{t=1}^{T}}$ (\ie, $T$ levels). 
Each codebook $C^t$ consists of $K$ dense code vectors $\mathcal{C}^t=\{c_k^t\}_{k=1}^{K}$ (\ie, $K$ latent factors), where $c_k^t \in \mathbb{R}^{D_q}$ and $D_q$ is the hidden dimension. 

First, we send the semantic vector $v$ into a deep neural network (DNN) encoder, and obtain the output vector $x \in \mathbb{R}^{D_q}$. 
At the first level ($t$=1), the residual vector is initialized as $r^1=x$. 
Then, for each quantization level $t$, the residual $r^t$ is quantized by mapping it to the nearest vector $c_{m^t}^t$ in codebook $\mathcal{C}^t$, where $m^t$ is the position index. The quantization at $t$-th level can be written as:
\begin{equation}
\begin{aligned}
    m^t &= \arg \min\nolimits_{k}
 \|r^t - c_k^t \|_2^2, \\
    r^{t+1} &= r^t - c_{m^t}^t.
\end{aligned}
\end{equation}
This process repeats recursively for $T$ times using codebooks ${\{\mathcal{C}^t\}_{t=1}^{T}}$. 
After obtaining the quantization indices $\mathcal{K} = \{m^t\}_{t=1}^{T}$, the quantized representation of $x$ can be acquired by $\hat{x} = \sum_{t=1}^{T} c_{m^t}^t$. 

Then $\hat{x}$ will be fed back into the DNN decoder. The output vector $\hat{v}$ of the decoder is used to reconstruct the semantic vector $v$. The training objective is defined as:
\begin{equation}
\begin{aligned}
    \mathcal{L}_{rec} &= \|v - \hat{v}\|_2^2, \\
    \mathcal{L}_{com} &= \sum\nolimits_{t=1}^{T} \left\| sg[r^t] - c_{m^t}^t \right\|_2^2 + \beta \left\| r^t - sg[c_{m^t}^t] \right\|_2^2, \\
    \mathcal{L}_{rq} &= \mathcal{L}_{rec} + \mathcal{L}_{com},
\end{aligned}
\end{equation}
where $sg[\cdot]$ is the stop-gradient operation and $\beta$ is the loss coefficient. 
The encoder, decoder and codebooks are jointly optimized by the loss $\mathcal{L}_{rq}$ which consists of two parts -- $\mathcal{L}_{rec}$ is the reconstruction loss, and $\mathcal{L}_{com}$ is the commitment loss that encourages residuals to stay close to the selected vectors in the codebooks. 
Due to the distinct semantic knowledge for user and item side, we train two sets of parameters for users and items separately. 
As a result, we quantize user and item semantic vectors (\ie,  $\{v^u_j\}_{j=1}^{|\mathcal{U}|}$ and $\{v^i_j\}_{j=1}^{|\mathcal{I}|}$) into latent factors $\mathcal{Q}^u = \{\mathcal{K}^u_j\}_{j=1}^{|\mathcal{U}|}$ and $\mathcal{Q}^i =  \{\mathcal{K}^i_j\}_{j=1}^{|\mathcal{I}|}$ respectively.


\subsubsection{Graph Construction} 

The extracted latent factors reflect the potential connections among users/items in various aspects, providing us an avenue to measure the global semantic relevance of users/items. 
Therefore, we incorporate the latent factors as extra nodes to equip the basic graph with a global view and in-depth semantics.
Specifically, the original node set $\mathcal{V} = \{\mathcal{U}, \mathcal{I}\}$ in Section~\ref{sec:preliminary} is extended to $\mathcal{V} = \{\mathcal{U}, \mathcal{I}, \mathcal{Q}^u, \mathcal{Q}^i\}$, with user and item latent factors $\mathcal{Q}^u$ and $\mathcal{Q}^i$ as additional nodes. Correspondingly, the edge set $\mathcal{E}$ consists of following types of edges:
\begin{itemize}[leftmargin=10pt]
    \item \textbf{User-Item Edge} $e_{u-i}$ connects the item $i$ and its positively interacted user $u$.
    \item \textbf{User-User Latent Factor Edge} $e_{u-q}$ connects each user node with his/her corresponding set of latent factor nodes $\mathcal{K}^u$ learned in Section~\ref{sec: Latent Factor Extraction}. The edges
    indicate the relationship that the multifaceted profile of each user can be semantically described by their $\mathcal{K}^u$.
    \item \textbf{Item-Item Latent Factor Edge} $e_{i-q}$ connects each item node with its corresponding set of latent factor nodes $\mathcal{K}^i$ learned.
    The edges represent the relationship that the multifaceted attributes of each item can be semantically characterized by their $\mathcal{K}^i$.
\end{itemize}


In this way, the one-hop neighbor nodes of items are composed of extracted latent factors and interacted users, providing semantic information and collaborative information respectively. With the shared latent factors as bridge, the two-hop items include more semantically similar ones, leading to a more informative neighborhood.  The user side is enhanced similarly. 

To be highlighted, the graph is automatically constructed based on the quantization learning process in Section~\ref{sec: Latent Factor Extraction}, rather than on predefined explicit relations. 
The learning process can establish the graph with a global view, which provides more comprehensive and informative insights through the topological structure. 
Meanwhile, we reduce the calls of LLMs to $O(N)$ complexity, which is more efficient compared with $O(N^2)$ complexity of existing LLM-based graph construction methods.



\subsection{Graph-enhanced Recommendation}
As shown in Figure~\ref{fig:framework}, since the constructed graph structure is heterogeneous and complex, we first define several metapaths to analyze the graph structure, and then design the \textit{metapath-based message propagation} for graph-enhanced user/item representations. 
Finally, the graph-enhanced representations can be integrated into downstream recommender systems in a model-agnostic manner, which is referred to \textit{recommendation enhancement}. 

\subsubsection{Metapath-based Message Propagation}
We aim to acquire the graph-enhanced user and item representations for the downstream recommendation tasks. 
To handle the heterogeneous and complex topological structure, we first define a set of metapaths to guide the message propagation on the graph. 
As depicted in Figure~\ref{fig:framework}, the metapath set $\mathcal{P}$ consists of the following four types:
\begin{itemize}[leftmargin=10pt]
    \item \textbf{Item-User Interaction Path} $i \rightarrow u$ means the collaborative information flow from an interacted item to the target user. 

    \item \textbf{User Semantic Path} $u \rightarrow q \rightarrow u$ indicates the semantic knowledge propagation between a pair of similar users who share the same latent factor node.

    \item \textbf{User-Item Interaction Path} $u \rightarrow i$ means the collaborative information flow from the target user to the interacted item. 

    \item \textbf{Item Semantic Path} $i \rightarrow q \rightarrow i$ indicates the semantic knowledge propagation between a pair of similar items that share the same latent factor node.
\end{itemize}
As shown in Figure~\ref{fig:framework}, based on these metapaths, we can build the subgraph for each user and item with their multi-hop neighbors. 
We denote the subgraph based on a certain type of metapath as $\{\mathcal{G}_p|p\in \mathcal{P}\}$. 
Then, we adopt graph attention networks (GATs)~\cite{velivckovic2017graph, wang2019kgat} as the message aggregator on these metapath-defined subgraphs.
The GAT operation for one target node $t$ is:
\begin{equation}
\begin{aligned}
\alpha_{tj} &= \frac{\exp\left(\operatorname{LeakyReLU}\left({\mathbf{a}}^T[W e_t \Vert W e_j]\right)\right)}{\sum_{k \in \mathcal{N}_t} \exp\left(\operatorname{LeakyReLU}\left({\mathbf{a}}^T[We_t \Vert W e_k]\right)\right)}, \\
h_t &= \sum\nolimits_{j \in \mathcal{N}_t} \alpha_{tj} W e_j, \\
\end{aligned}
\label{eq: original gat}
\end{equation}
where $e_t, e_k, e_j \in \mathbb{R}^{D_e}, h_t \in \mathbb{R}^{D_h}$ denote original and GAT-enhanced node embeddings respectively. $\mathcal{N}_t$ is the neighbor set of target node $t$. $W \in \mathbb{R}^{D_{h} \times D_e}$ is a learnable matrix, and $\mathbf{a}\in \mathbb{R}^{2D_h}$ is a trainable vector to compute attention logits. $\Vert$ denotes the vector concatenation operation.




We apply the GAT operation in Equation~\ref{eq: original gat} sequentially based on the pre-defined metapaths, and obtain the graph representations for the target user/item nodes. 
Let $\mathbf{E}=\{e^u, e^i, e^{Q_u}, e^{Q_i}\}$ denote the trainable node embeddings of users, items, user latent factors and item latent factors, respectively. 
We first apply GAT based on two semantic metapaths (\ie, $u \rightarrow q \rightarrow u$ and $i \rightarrow q \rightarrow i$) to obtain the semantically enhanced representations for users and items:
\begin{equation}
\begin{aligned}
\mathcal{H}^u &= \operatorname{GAT}(\mathbf{E}, \mathcal{G}_{u \rightarrow q \rightarrow u}), \\
\mathcal{H}^i &= \operatorname{GAT}(\mathbf{E}, \mathcal{G}_{i \rightarrow q \rightarrow i}),
\end{aligned}
\end{equation}
where $\mathcal{H}^u$ and $\mathcal{H}^i$ aggregate the semantic knowledge with the latent factors as the bridge. 
Then, we further model the user-item collaborative information based on the interaction metapaths (\ie, $u \rightarrow i$ and $i\rightarrow u$) to acquire the final graph-enhanced representations of target user/items:
\begin{equation}
\begin{aligned}
\hat{\mathcal{H}}^u &= \operatorname{GAT}(\mathcal{H}^u\cup \mathcal{H}^i ,  \mathcal{G}_{i \rightarrow u}), \\
\hat{\mathcal{H}}^i &= \operatorname{GAT}(\mathcal{H}^u\cup \mathcal{H}^i, \mathcal{G}_{u \rightarrow i}).
\end{aligned}
\end{equation}

With the metapath-based message propagation, we are able to fully fuse the in-depth semantic knowledge deduced by LLMs and the collaborative information based on interaction records, resulting in graph-enhanced user/item representations for downstream recommendation enhancement.

\subsubsection{Recommendation Enhancement}
Since \modelname\ is a model-agnostic framework, the obtained graph-enhanced user/item representations can be integrated into arbitrary downstream recommender systems in various ways. 
In this paper, we simply integrate them as auxiliary features to improve the preference estimation of a user $u$ towards a target item $i$:
\begin{equation}
\begin{aligned}
score &= \Phi (\mathcal{S}^u, \mathcal{F}^u, \mathcal{F}^i, \hat{\mathcal{H}}^u, \hat{\mathcal{H}}^i).
\end{aligned}
\end{equation}


\subsection{More Discussions}
We provide further discussions about AutoGraph to address readers' possible concerns: (1) What is the difference between the graph constructed by AutoGraph and knowledge graphs?
(2) How can residual quantization equip the graph with a global view of in-depth semantics? 
(3) How can AutoGraph be industrially deployed? Due to the page limitation, we provide the discussion in Appendix~\ref{app: discussions}.

%% file: text/experiment.tex
\section{Experiment}


\subsection{Experiment Setup}

\subsubsection{Datasets}
We conduct experiments on three public datasets, \ie, MovieLens-1M, Amazon-Books and BookCrossing. Due to the page limitation,
we show the dataset statistics and give detailed data preprocessing information in Appendix~\ref{app: data preprocess}. 

\subsubsection{Evaluation Metrics}
Following previous works~\cite{he2015trirank, xi2023device, he2017neural}, we adopt four widely used metrics, \ie, top-$K$ Normalized Discounted Cumulative Gain (NDCG@K), top-$K$ Hit Ratio (HR@K), Mean Reciprocal Rank (MRR) and Group Area under the ROC curve (GAUC). 
Higher values of these metrics indicate better recommendation performance. 
The truncation level $K$ is set to 10.

\subsubsection{Backbone Models} 
As a model-agnostic framework, \modelname~can be incorporated with various recommendation models by providing the graph-enhanced representations. 
In this paper, we select four representative models as backbones to validate the effectiveness of \modelname\ for recommendation, \ie, \textbf{YouTubeDNN}~\cite{covington2016deep}, \textbf{MIND}~\cite{li2019multi}, \textbf{GRU4Rec}~\cite{hidasi2015session} and \textbf{SASRec}~\cite{kang2018self}. They generally cover four different core operators for user behavior modeling~\cite{he2023survey}: deep neural networks (DNNs), capsule networks~\cite{sabour2017dynamic}, gated recurrent units (GRUs)~\cite{chung2014empirical} and self-attention mechanisms~\cite{vaswani2017attention} respectively.


\begin{table*}
\vspace{-10pt}
\caption{The performance of \modelname\ and baseline methods based on different backbone models. ``*+'' denotes the backbone is enhanced by certain baseline method. 
The best result is in bold, and the second-best value is underlined. 
\textit{Rel.Impr} denotes the relative improvement of our \modelname\ framework against the best baseline result. 
The symbol * indicates statistically significant improvement over the best baseline with $p$-value < 0.001. 
\textit{NG} is short for \textit{NDCG}, and \textit{HR} is short for \textit{Hit Ratio}.
}
\vspace{-10pt}
\label{tab:main table}
\resizebox{0.95\textwidth}{!}{
\renewcommand\arraystretch{1.1}
\begin{tabular}{l|cccc|cccc|cccc}
\toprule
\hline

\multirow{2}{*}{Model} & \multicolumn{4}{c|}{MovieLens-1M} & \multicolumn{4}{c|}{Amazon-Books} & \multicolumn{4}{c}{BookCrossing} \\ 
 & NG@10 & HR@10 & MRR & GAUC & NG@10 & HR@10 & MRR & GAUC & NG@10 & HR@10 & MRR & GAUC \\ 
\hline
\textbf{YouTubeDNN}  & 0.0418 & 0.0888 & 0.0402 & 0.8947 & 0.0432 & 0.0895 & 0.0394 & 0.7996 & 0.0845 & 0.1388 & 0.0779 & 0.7509 \\
*+KAR & \underline{0.0492} & 0.1015 & \underline{0.0463} & \underline{0.9114} & 0.0508 & 0.0987 & 0.0467 &  0.8053  & 0.0873 & 0.1493 & 0.0794 & 0.7745 \\
*+UIST  & 0.0467 & 0.0956 & 0.0452 & 0.9076 & 0.0514 & \underline{0.1017} & 0.0470 & 0.8102 & 0.0906 & 0.1515 & 0.0837 & \underline{0.7793} \\
*+LightGCN  & 0.0426 & 0.0906 & 0.0405 & 0.8963 & 0.0444 & 0.0845 & 0.0410 &  \textbf{0.8417} & 0.0832 & 0.1394 & 0.0773 & 0.7517\\
*+CCGNN  & 0.0489 & \underline{0.1024} & 0.0457 & 0.8957 & \underline{0.0517} & 0.0974 & \underline{0.0478} & 0.7968 & \underline{0.0922} & \underline{0.1525} & \underline{0.0845}  & 0.7754\\
*+TopoI2I  & 0.0468 & 0.0994 & 0.0440 & 0.9010 & 0.0491 & 0.0945 & 0.0458  & 0.7984 & 0.0894 & 0.1478 & 0.0824 & 0.7714 \\
*+\modelname  & \textbf{0.0707$^*$} & \textbf{0.1221$^*$} & \textbf{0.0686$^*$} & \textbf{0.9128$^*$} & \textbf{0.0609$^*$} & \textbf{0.1173$^*$} & \textbf{0.0545$^*$} &  \underline{0.8362} & \textbf{0.0948$^*$} & \textbf{0.1560$^*$} & \textbf{0.0869$^*$} & \textbf{0.7811$^*$}\\
Rel.Impr  & 43.57\% & 19.28\% & 48.14\% & 0.15\% & 17.74\% & 15.38\% & 14.13\%  &  -0.65\% & 2.87\% & 2.29\% & 2.79\% & 0.23\%\\
\hline
\textbf{MIND}  & 0.0462 & 0.0979 & 0.0437 & 0.8932 & 0.0519 & 0.1010 & 0.0461 & 0.7993 & 0.0851 & 0.1395 & 0.0797 & 0.7793\\
*+KAR  & 0.0505 & 0.1007 & 0.0485 & 0.8990 & 0.0567 & 0.1113 & 0.0509 & 0.8041 & 0.0862 & 0.1469 & 0.0788 & 0.7914\\
*+UIST  & 0.0500 & 0.1037 & 0.0471 & \underline{0.9035} & \underline{0.0579} & \underline{0.1122} & 0.0518 & \underline{0.8246} & 0.0931 & \underline{0.1573} & 0.0843 & \underline{0.8029}\\
*+LightGCN  & 0.0435 & 0.0931 & 0.0410 & 0.8936 & 0.0532 & 0.1029 & 0.0487 & 0.8077 & 0.0853 & 0.1418 & 0.0796 & 0.7803\\
*+CCGNN  & \underline{0.0543} & \underline{0.1136} & \underline{0.0514} & 0.9034 & 0.0535 & 0.1063 & 0.0486 & 0.8169 & \underline{0.0938} & 0.1556 & \underline{0.0860} & 0.7946\\
*+TopoI2I  & 0.0514 & 0.1050 & 0.0486 & 0.9021 & 0.0577 & 0.1117 & \underline{0.0521} & 0.8199 & 0.0907 & 0.1499 & 0.0839 & 0.7824\\
*+\modelname  & \textbf{0.0643$^*$} & \textbf{0.1166$^*$} & \textbf{0.0622$^*$} & \textbf{0.9126$^*$} & \textbf{0.0737$^*$} & \textbf{0.1373$^*$} & \textbf{0.0656$^*$} & \textbf{0.8297$^*$} & \textbf{0.0989$^*$} & \textbf{0.1687$^*$} & \textbf{0.0897$^*$} & \textbf{0.8114$^*$}\\
Rel.Impr  & 18.41\% & 2.62\% & 21.05\% & 1.01\%  & 27.32\% & 22.39\% & 26.07\% & 0.62\% & 5.46\% & 7.24\% & 4.24\% & 1.06\%\\
\hline
\textbf{GRU4Rec}  & 0.0812 & 0.1586 & 0.0730 & 0.9200 & 0.0754 & 0.1466 & 0.0653  & 0.8371 & 0.0969 & 0.1636 & 0.0880 & 0.8002\\
*+KAR  & \underline{0.0903} & \underline{0.1750} & 0.0786 & 0.9245 & 0.0824 & 0.1491 & 0.0742 & 0.8466 & 0.0986 & 0.1579 & 0.0917 & 0.7914\\
*+UIST & 0.0889 & 0.1742 & 0.0771 & 0.9233 & \underline{0.0901} & \underline{0.1644} & \underline{0.0791} & 0.8382   & 0.1062 & 0.1743 & 0.0968 & 0.8052\\
*+LightGCN  & 0.0837 & 0.1569 & 0.0734 & 0.9177 & 0.0778 & 0.1458 & 0.0689 & 0.8272 & 0.0958 & 0.1613 & 0.0872 & 0.7965\\
*+CCGNN  & 0.0862 & 0.1679 & 0.0774 & 0.9248 & 0.0864 & 0.1608 & 0.0753 & 0.8300 & \underline{0.1063} & \underline{0.1748} & \underline{0.0974} & 0.8162\\
*+TopoI2I  & 0.0891 & 0.1745 & \underline{0.0788} & \underline{0.9249} & 0.0814 & 0.1556 & 0.0706 & \underline{0.8502} & 0.1029 & 0.1681 & 0.0944 & \underline{0.8175}\\
*+\modelname  & \textbf{0.0937$^*$} & \textbf{0.1790$^*$} & \textbf{0.0854$^*$} & \textbf{0.9291$^*$} & \textbf{0.0959$^*$} & \textbf{0.1761$^*$} & \textbf{0.0837$^*$} & \textbf{0.8553$^*$} & \textbf{0.1093$^*$} & \textbf{0.1801$^*$} & \textbf{0.1002$^*$} & \textbf{0.8265$^*$}\\
Rel.Impr  & 3.79\% & 2.29\% & 8.44\% & 0.45\% & 6.38\% & 7.09\% & 5.84\% & 0.60\% & 2.80\% & 3.01\% & 2.90\% & 1.10\%\\
\hline
\textbf{SASRec}  & 0.0823 & 0.1672 & 0.0721 & 0.9245 & 0.0802 & 0.1543 & 0.0704 & 0.8470 & 0.0952 & 0.1585 & 0.0878 & 0.8142\\
*+KAR  & 0.0896 & 0.1788 & 0.0786 & \underline{0.9265} & 0.0852 & 0.1588 & 0.0748 & \underline{0.8666} & 0.1043 & 0.1760 & 0.0944 & 0.8138\\
*+UIST  & 0.0882 & 0.1707 & 0.0787 & 0.9258 & 0.0862 & \underline{0.1638} & 0.0751 & 0.8630 & \underline{0.1078} & \underline{0.1774} & \underline{0.0982} & \underline{0.8298}\\
*+LightGCN & 0.0829 & 0.1652 & 0.0733 & 0.9250 & 0.0845 & 0.1553 & 0.0750 &0.8554  & 0.0960 & 0.1596 & 0.0882 & 0.8145\\
*+CCGNN  & 0.0867 & 0.1717 & 0.0768 & 0.9258 & 0.0855 & 0.1588 & 0.0737 & 0.8463 & 0.1055 & 0.1769 & 0.0951 & 0.8230\\
*+TopoI2I  & \underline{0.0917} & \underline{0.1825} & \underline{0.0798} & 0.9262 & \underline{0.0864} & 0.1604 & \underline{0.0759} & 0.8637 & 0.1026 & 0.1684 & 0.0945 & 0.8177\\
*+\modelname  & \textbf{0.1047$^*$} & \textbf{0.1924$^*$} & \textbf{0.0941$^*$} & \textbf{0.9330$^*$} & \textbf{0.0959$^*$} & \textbf{0.1797$^*$} & \textbf{0.0824$^*$} & \textbf{0.8722$^*$} & \textbf{0.1114$^*$} & \textbf{0.1852$^*$} & \textbf{0.1011$^*$} & \textbf{0.8376$^*$}\\
Rel.Impr  & 14.22\% & 5.46\% & 17.98\% & 0.70\% & 10.99\% & 9.76\% & 8.55\% & 0.65\% & 3.28\% & 4.36\% & 2.97\% & 0.94\% \\
\hline
\bottomrule
\end{tabular}
\vspace{-20pt}
}
\end{table*}

\begin{table*}
\vspace{-7pt}
\caption{Comparison of graph construction efficiency of different methods. $N$ denotes the number of items and $E$ denotes the number of sampled candidates for each item to compute similarity with. \textit{Avg. GAUC Rel.Impr} is the average GAUC relative improvement of different graph construction methods over the four backbone models on the three datasets .
}
\vspace{-10pt}
\label{tab:efficiency}
\resizebox{0.85\textwidth}{!}{
\renewcommand\arraystretch{1.1}
\begin{tabular}{c|cccc|cccc|c}
\toprule
\hline
\multirow{3}{*}{\makecell{Graph Construction \\ Methods}} & \multicolumn{4}{c|}{Initial Graph Construction} & \multicolumn{4}{c|}{Incremental Node Insertion} & \multirow{3}{*}{\makecell{Avg. GAUC \\Rel.Impr}} \\ \cline{2-9}
 & \multicolumn{1}{c|}{\multirow{2}{*}{\makecell{Time \\ Complexity}}} & \multicolumn{3}{c|}{Run Time} & \multicolumn{1}{c|}{\multirow{2}{*}{\makecell{Time \\ Complexity}}} & \multicolumn{3}{c|}{Run Time} &  \\
 & \multicolumn{1}{c|}{}  & ML-1M & Amz-Books & BX & \multicolumn{1}{c|}{}  & ML-1M & Amz-Books & BX &  \\ \hline
LightGCN & \multicolumn{1}{c|}{$O(1)$}  & 1.75s & 4.26s & 59.13s & \multicolumn{1}{c|}{$N/A$} & $N/A$ & $N/A$ & $N/A$ & 0.50\% \\
CCGNN & \multicolumn{1}{c|}{$O(N)$}  & 5min & 14min & 3h & \multicolumn{1}{c|}{$O(1)$} & 0.085s & 0.103s & 0.084s & 0.93\% \\
TopoI2I & \multicolumn{1}{c|}{$O(NE)$} & 190h & 315h & 4700h & \multicolumn{1}{c|}{$O(E)$}  & 193.603s & 132.911s & 131.144s & 1.18\% \\
\modelname & \multicolumn{1}{c|}{$O(N)$}  & 18h & 26h & 202h & \multicolumn{1}{c|}{$O(1)$}  & 6.640s & 6.637s & 5.805s & 2.81\% \\ \hline
\bottomrule
\end{tabular}
\vspace{-25pt}
}

\end{table*}

\subsubsection{Baseline Methods}

Based on the backbone models, various model-agnostic methods could be applied to promote  recommendation performance.
Since \modelname\ harnesses LLMs for automatic graph construction and enhancement, we select the baseline methods from two perspectives: 
(1) \textit{LLM-augmented methods} that leverage the open-world knowledge and reasoning abilities of LLMs to enhance the performance. 
\textbf{KAR}~\cite{xi2023towards} designs specific prompts to extract user/item knowledge from LLMs, which is further encoded as additional features for recommendation models. 
\textbf{UIST}~\cite{liu2024discrete} adopts the LLM-based discrete semantic tokens from vector quantization as auxiliary feature IDs. 
(2) \textit{Graph-enhanced methods} that construct and incorporate different types of graphs for recommendation enhancement. 
\textbf{LightGCN}~\cite{he2020lightgcn} conducts collaborative filtering based on the vanilla user-item interaction graph. \textbf{CCGNN}~\cite{xv2023commerce} uses NER techniques to extract phrases from item texts and incorporates them as explicit nodes and linkages for graph construction. 
\textbf{TopoI2I}~\cite{sun2023large} takes large language models as topological structure enhancers to deduce the pairwise item similarity for edge addition and removal, resulting in automatic graph construction for recommendation.



\subsubsection{Implementation Details}
\label{sec: implementation}

We provide implementation details for \modelname\ and baselines in Appendix~\ref{app: implementation} due to page limitation. Our code is available\footnote{\url{https://github.com/LaVieEnRose365/AutoGraph}}.

\subsection{Overall Performance}
We evaluate the performance of \modelname\ based on different backbone models in comparison to existing baseline methods. The results are reported in Table~\ref{tab:main table}, from which we can have the following observations:
\begin{itemize}[leftmargin=10pt]
    \item  \modelname\ is model-agnostic and highly compatible with various backbone models. The information from our constructed graph is supplementary to the backbone models, significantly enhancing recommendation performance across them.

    \item \modelname\ surpasses \textit{LLM-augmented baseline methods} on all three datasets. Although these methods (\ie, KAR and UIST) leverage the semantic knowledge of LLMs, they fail to utilize the multi-hop neighbor information based on the semantics. In contrast, \modelname\ explores the relationships between semantically relevant nodes and integrates the multi-hop neighbor information, resulting in better performance.


    \item \modelname\ generally outperforms existing \textit{graph-enhanced baseline methods} by a significant margin. LightGCN and CCGNN employ specific rules and fail to model the complex semantic signals, while TopoI2I focuses on local-view pairwise comparison and is unable to capture the high-quality topological structure provided by the global view. 
    In comparison, based on the latent factors from LLM knowledge, \modelname\ not only captures in-depth semantics in multiple facets, but also equips the graph with a global view, leading to superior performance. 

\end{itemize}

\subsection{Efficiency Comparison}
We analyze the graph construction efficiency of \modelname\ and different graph-enhanced methods here. Specifically, we compare the time complexity and real running time in two different cases:
\begin{itemize}[leftmargin=10pt]
    \item \textit{Initial graph construction} refers to constructing the graph from scratch based on existing user profiles and item attributes.
    \item \textit{Incremental node insertion} refers to the phase when a new user/item is introduced and needs to be added into the constructed graphs.
\end{itemize}
The results are reported in Table~\ref{tab:efficiency}. Due to page limitation, we explain more details of evaluation settings in Appendix~\ref{app: efficiency}. Next we provide a comprehensive evaluation of different graph construction methods:
\begin{itemize}[leftmargin=10pt]
    \item Since LightGCN constructs graphs based on simple rules (\eg, click as linkage), it costs least time but performs worst, as it is vulnerable to noisy clicks and fails to harness the semantic information. Moreover, it falls short in the case of \textit{incremental node insertion} and is unable to adapt to the dynamic and fast-evolving data in industrial scenarios.
    \item Both CCGNN and TopoI2I explore the semantic information and perform better than LightGCN. Since TopoI2I leverages LLMs to better enrich the semantics, it performs relatively better than CCGNN. However, the pairwise comparison nature of TopoI2I induces more computation overhead. Even if TopoI2I applies pre-filtering to downsample the LLM invocations, the efficiency and real running time are still poor.
    \item AutoGraph is the most cost-effective compared to baselines, showing superior performance in terms of both efficacy and efficiency. AutoGraph incorporates the in-depth semantics from LLMs and equips the constructed graph with a global view, thus leading to better efficacy. Besides, AutoGraph reduces the calls of LLMs to $O(N)$ complexity, resulting in better efficiency. 

\end{itemize}

\subsection{Ablation Study}


We investigate on the impact of different configurations and hyperparameters in \modelname. Due to the page limitation, here we only show experiments on contribution of different metapaths to the graph. More other experiments are shown in Appendix~\ref{sec: more experiments}.

As is shown in Table~\ref{tab:ablation study}, we remove different metapaths from the subgraphs to evaluate their efficacy respectively, \ie, user semantic path ($u \rightarrow q \rightarrow u$), item semantic path ($i \rightarrow q \rightarrow i$), and interaction paths ($u \rightarrow i$ and $i \rightarrow u$). Moreover, $N/A$ means removing all the metapaths, \ie, the vanilla backbone model. 
 
 Removing different metapaths serves as the ablation study w.r.t. the item/user representation enhancement brought by AutoGraph. We can observe that removing each metapath of \modelname\ generally results in performance degradation, while these variants still outperform the vanilla backbone models. This demonstrates the efficacy of each metapath proposed in our 
 \modelname\ framework. Moreover, the semantic paths contribute more to the performance improvement than the interaction paths, highlighting the importance of in-depth semantics from LLMs and global-view information brought by quantization-based latent factors.

\begin{table}[t]
\caption{
Ablation study w.r.t. different metapaths. We remove different metapaths from the subgraphs of \modelname\ to evaluate their contribution respectively. $N/A$ means the vanilla backbone. The best result is given in bold, and the second-best value is underlined.
}
    \vspace{-10pt}
\label{tab:ablation study}
\resizebox{\linewidth}{!}{
\renewcommand\arraystretch{1.1}
\begin{tabular}{c|c|cccc|cccc}
\toprule
\hline
 \multicolumn{1}{c|}{\multirow{2}{*}{\makecell{Backbone\\ Model}}} & \multicolumn{1}{c|}{\multirow{2}{*}{Graph Variants}}  &  \multicolumn{4}{c|}{MovieLens-1M} &  \multicolumn{4}{c}{Amazon-Books} \\ 
 \multicolumn{1}{c|}{} & \multicolumn{1}{c|}{} &  NG@10 & HR@10 & MRR & GAUC & NG@10 & HR@10 & MRR & GAUC \\ \hline
 
\multicolumn{1}{c|}{\multirow{5}{*}{YT-DNN}} & \modelname\ (Ours) & \textbf{0.0707} & \textbf{0.1221} & \textbf{0.0686} & \textbf{0.9128} & \textbf{0.0609} & \textbf{0.1173} & \textbf{0.0545} & \textbf{0.8362}       \\
& w/o $u\rightarrow q \rightarrow u$ & 0.0526 & 0.0990 & 0.0516 & 0.9054 & 0.0491 & 0.0920 & 0.0458 & 0.8021 \\
& w/o $i\rightarrow q\rightarrow i$ & 0.0454 & 0.0961 & 0.0434 & 0.9008 & 0.0480 & 0.0974 & 0.0433 & 0.7986 \\
& w/o $u\rightarrow i$ \& $i\rightarrow u$  & \underline{0.0546} & \underline{0.1065} & \underline{0.0522} & \underline{0.9061} & \underline{0.0546}  & \underline{0.1055} & \underline{0.0492} & \underline{0.8148} \\
& $N/A$  & 0.0418 & 0.0888 & 0.0402 & 0.8947 & 0.0432 & 0.0895 & 0.0394 & 0.7996 \\
 \hline

\multicolumn{1}{c|}{\multirow{5}{*}{MIND}} & \modelname\ (Ours) & \textbf{0.0643} & \textbf{0.1166} & \textbf{0.0622} & \textbf{0.9126} & \textbf{0.0737} & \textbf{0.1373} & \textbf{0.0656} & 0.8297 \\
& w/o $u\rightarrow q \rightarrow u$ & 0.0560 & 0.1057 & 0.0545 & \underline{0.9120} & 0.0593 & 0.1163 & 0.0532 & 0.8192  \\
& w/o $i\rightarrow q\rightarrow i$ & 0.0467 & 0.0975 & 0.0450 & 0.9003 & 0.0553 & 0.1058 & 0.0504 & \underline{0.8343} \\
& w/o $u\rightarrow i$ \& $i\rightarrow u$   & \underline{0.0594} & \underline{0.1102} & \underline{0.0572} & 0.9102 & \underline{0.0662} & \underline{0.1263} & \underline{0.0590} & \textbf{0.8344}  \\
& $N/A$  & 0.0462 & 0.0979 & 0.0437 & 0.8932 & 0.0519 & 0.1010 & 0.0461 & 0.7993 \\
 \hline

 \bottomrule 
\end{tabular}
}
\vspace{-10pt}
\end{table}

\subsection{Industrial Deployment}
To evaluate the performance of AutoGraph, we conduct an online A/B test in Huawei's online advertising platform for ten consecutive days, where hundreds of millions of impressions are generated these days. Specifically, 10\% of users are randomly allocated to the experimental group, and another 10\% to the control group. For the control group, the users are served by a highly optimized deep model. For the experimental group, the users are served by the same base model with AutoGraph. We utilize Huawei’s large language model PanGu~\cite{zeng2021pangu} to generate user and item knowledge, and assist the recommendation with graph-enhanced representations for the experimental group. We offer more details and suggestions of industrial deployment of our AutoGraph framework in Appendix~\ref{app: industrial deployment}. 

We compare the performance according to two metrics: RPM (Revenue Per Mille), and eCPM (Effective Cost Per Mille), which are widely used for online advertising to measure online revenue~\cite{yang2024aie,ecpm}. In the online A/B test, AutoGraph achieves 2.69\% improvements on RPM and 7.31\% improvements on eCPM over the base model. It is a significant improvement and sufficiently validates the effectiveness of our model in industrial applications. After 2 weeks
of evaluation, AutoGraph has become the main model in this scenario to carry most of the online traffic.



%% file: text/related_work.tex
\section{Related Work}
\label{sec: related work}

\subsection{Graph-enhanced Recommendation}
Over the past decade, GNN-based recommender systems~\cite{ying2018graph, he2020lightgcn, de2024personalized} have become new state-of-the-art due to the power of graphs in capturing relationships~\cite{grover2016node2vec, xu2018powerful, hamilton2017inductive, perozzi2014deepwalk}. Most of the existing graph-enhanced recommendation methods focus on the optimization of model structures~\cite{shi2018heterogeneous,wang2023knowledge} and improvement of learning strategies~\cite{jiang2023adaptive,wu2021self} based on pre-defined graphs, while the importance of the graph construction stage is generally overlooked. Typically, earlier works usually construct graphs based on specific rules. Most of them construct a bipartite graph of users and items with click as linkage~\cite{he2020lightgcn, jin2020multi, chang2021sequential}. 
Other works~\cite{xv2023commerce, wang2019knowledge, wang2019kgat, lin2015learning, wang2018ripplenet, ji2021survey} further incorporate the semantic relationships of users and items. 
However, these designed rules (\eg, click as linkage, and entity extraction~\cite{xv2023commerce}) can only explore the superficial semantic relationships and fall short of modeling the sophisticated semantic signals in recommendation data.  Besides, there is also a line of works focusing on relational annotations (\eg, knowledge graphs)~\cite{wang2019knowledge, wang2019kgat, ji2021survey}. The annotations usually require significant human resources and specialized expertise, which is impractical in large-scale industrial scenarios. As a result, it is meaningful to design an automatic graph construction framework which can explore deep semantics and is efficient for large-scale industrial applications.


\subsection{LLMs for Graph Construction}
While large language models have been applied in numerous complex practical tasks and showcase significant potential~\cite{bao2023tallrec, jiang2024survey, ahn2024large, lin2023can,lin2024clickprompt, damianou2024towards}, research on LLMs for graph topology enhancement is still at an early stage~\cite{xu2024multi, zhang2023making, shang2024survey, pan2024unifying}, and there is large blank especially for LLM-based graph topology enhancement in recommender systems. Outside the field of recommender systems, the paradigm for LLM-based graph topology enhancement focuses on pairwise similarities of nodes. Specifically, LLMs are prompted to directly infer the similarity score between two nodes with in-context learning strategy~\cite{sun2023large, brown2020language} or instruction tuning strategy~\cite{guo2024graphedit}. Then edges will be added or deleted based on the deduced similarities. Other works mainly lie in LLMs for Knowledge Graph Completion~\cite{chen2020knowledge, hsu2021retagnn, zhang2023making, xu2024multi}. They focus on discovering the possible relations of two entities, still belonging to the pairwise comparison paradigm.

However, the pairwise paradigm has disadvantages in both effectiveness and efficiency in recommendation. First, the global view of the entire dataset (\eg, contextual information) is crucial for performance improvement~\cite{yang2024give}, while the pairwise comparison is limited to local topology refinement of graphs and can hardly build a comprehensive global view, due to the large scale of recommendation data and context window limitation of LLMs.
Second, the pairwise comparison of nodes incurs a time complexity of $O(N^2)$, while the users, items and their attributes in recommendation easily reach the scale of millions~\cite{li2019multi}, making it impractical to be applied in real recommender systems.

In this paper, we mainly focus on how to effectively and efficiently construct graphs with LLMs on large-scale recommendation data for industrial applications. To the best of our knowledge, we are the first to harness LLMs and vector quantization for graph construction with a global view in recommendation. A novel \modelname\ framework is proposed to enhance the graph structure with a global semantic insight, and demonstrates both effectiveness and efficiency in contrary to the existing pairwise comparison paradigm of LLMs for enhanced graph construction.

%% file: text/conclusion.tex
\section{Conclusion}
In this paper, we propose a novel framework (\ie, AutoGraph) for automatic graph construction based on LLMs for recommendation. 
We extract the latent factors of LLM-enhanced user/item semantic vectors based on quantization techniques. 
The process reduces the calls of LLMs to $O(N)$ complexity and improves efficiency over existing LLM-based graph construction methods. 
With the latent factors as extra nodes, the constructed graph can not only fully extract the in-depth semantics, but also establish a global view. Furthermore, we design metapath-based message propagation to effectively aggregate the semantic and collaborative information. The framework is model-agnostic and compatible with different recommender systems. Extensive experiments on three real-world datasets validate the efficacy and efficiency of \modelname\ compared with baseline models. We have deployed AutoGraph in Huawei advertising
platform and gained improvement in the online A/B test.  Up to now, AutoGraph has been the main model to carry out the major traffic in this scenario.

%% file: text/appendix.tex
\newpage

\appendix

\section{More Discussions}
\label{app: discussions}
In this section, we provide more discussions about our proposed AutoGraph framework to address readers' possible questions, \ie, (1) the difference between the graph constructed by AutoGraph and knowledge graphs, (2) how residual quantization can equip the graph with a global view of in-depth semantics, and (3) how AutoGraph can be industrially deployed.

\subsection{Difference with Knowledge Graphs}
In this paper, we propose a novel graph construction framework (\ie, AutoGraph) that is different from knowledge graphs (KGs) in following aspects:
\begin{itemize}[leftmargin=10pt]
    \item \textbf{Single side \textit{v.s.} Dual side}. Knowledge graphs are usually established at the item side, failing to enhancing the user side. However, the user information is indispensable to recommender systems and enhancing the user-side representations is important to recommendation improvement. In comparison, \modelname\ enhances both the item side and user side, thus making both item and user representations more informative, leading to better performance. 

    \item \textbf{Explicit entities \textit{v.s.} Implicit concepts}. Extra nodes introduced in knowledge graphs are usually explicit named entities, which can only explore shallow semantics. 
    In comparison, \modelname\ learns the implicit concepts that encode a distribution of latent factors based on LLMs, which helps extract in-depth and sophisticated semantics, thus improving the recommendation performance. 

    \item \textbf{Predefined relations \textit{v.s.} Automatic construction}. The edges for entity node linkage in knowledge graphs are manually defined relations, while AutoGraph does not require the explicit relation definition and automates the graph construction based on LLMs and vector quantization. 
    This makes AutoGraph more flexible and scalable than KGs.
\end{itemize}

\subsection{Global View with Quantization}
In our AutoGraph framework, we propose to leverage quantization techniques for graph construction with a global view of in-depth semantics. Specifically, quantization equips our graph with a global view in following ways:
\begin{itemize}[leftmargin=10pt]
    \item Connections between latent factor nodes and user/item nodes can have mutual effects on each other. Linking a target user/item to a certain latent factor not only locally influences the user's/item's own representation, but also has broader global effects on other users or items connected by the assigned latent factor.

    \item The residual quantization iteratively approximates the user/item representation residuals. The coarse-to-fine manner structures the node neighbors hierarchically, integrating both broad global patterns and subtle local similarities.
\end{itemize}

\subsection{Industrial Deployment}
\label{app: industrial deployment}
The new introduced nodes of AutoGraph over the vanilla user-item graphs are the latent factor nodes, whose numbers are equal to the number of quantization codebook vectors. In our practice, thousand-level codebook vectors are enough to explore meaningful semantics. The number of new nodes is much smaller than the size of users and items, which easily reaches million level. And the number of edges is controllable by neighborhood sampling, leading to graphs of appropriate sizes. Next, we provide more details of our deployment strategy.

We mainly leverage AutoGraph as a plug-and-play graph-enhanced representation generator, which is compatible with existing recommender system architectures through an offline pre-storage strategy. The core process can be divided into three stages:
\begin{itemize}[leftmargin=10pt]
    \item \textbf{Existing Node Processing}: Offline pre-store LLM-enhanced semantic vectors, train multi-level quantization codebooks to construct enhanced graphs, and synchronize the pre-stored graph-enhanced representations for online use.
    \item \textbf{Incremental Node Processing}: For newly added/updated nodes, a single call to the LLM is made to obtain the semantic vector. A codebook nearest neighbor search is used to quickly assign latent factors and generate pre-stored graph-enhanced representations.
    \item \textbf{Fast-slow Model Update}: The main recommendation model is frequently refreshed, while we do not need to frequently re-train the residual quantization model (\ie, encoder, decoder and codebooks) since it is expressive and generalized. We suggest that the quantization model can be updated in fixed intervals (\eg, one week) to re-fit the evolving user/item distribution, while during the interval the latest one is used.
\end{itemize}
This solution maintains the efficiency of the online service by only adding negligible read overhead for pre-stored representations, avoiding the burden of real-time graph computation. By using offline asynchronous processing and periodic model updates, it strikes a balance between representation quality and system performance.

\section{data preprocessing}
\label{app: data preprocess}


For MovieLens-1M and Amazon-Books datasets, we only retain the interactions with rating above 3 as positive interactions, while for BookCrossing dataset we retain the interactions with rating above 5.The records are sorted according to timestamp and we filter out users whose behavior history length is less than 5. Following previous works~\cite{wang2019neural, qin2021retrieval, yang2024sequential}, each dataset is split using \textit{leave-one-out} strategy, with the last item for testing, the second-to-last item for validation and the remaining interactions for training. For training, we rank the ground-truth next item against 50 randomly sampled negative items. For testing, we rank the ground-truth item against the full item sets on the MovieLens-1M and Amazon-Books dataset. On the BookCrossing dataset, we follow previous works~\cite{wang2023sequential, guo2024integrating, nguyen2024towards} and randomly sample 1,000 negative items, as the full item scale is large. The maximum length of behavior sequence is set to 30. The dataset statistics are shown in Table~\ref{tab:datasets}.
\begin{table}[h]
    \vspace{-5pt}
    \caption{The datasets statistics.}
    \vspace{-10pt}
    \centering
    \resizebox{0.47\textwidth}{!}{
    \renewcommand\arraystretch{1.1}
    \begin{tabular}{c|ccccccc}
    \toprule
     Dataset   & \#Users & \#Items & \#Samples & \# User Features (Fields) &\# Item Features (Fields) \\ 
     \midrule
     \href{http://www2.informatik.uni-freiburg.de/~cziegler/BX/}{BookCrossing}  & 6,853 & 129,018 & 190,825 & 61,631 (3) & 310,719 (5) \\
     \href{https://grouplens.org/datasets/movielens/1m/}{MovieLens-1M} & 6,038 & 3,533 & 545,114 & 9,506 (5) & 7,084 (3) \\
     \href{https://cseweb.ucsd.edu/~jmcauley/datasets/amazon_v2/}{Amazon-Books} & 6,010 & 8,531 & 77,1325 &  6,010 (1) & 10,629 (3) \\ \bottomrule
    \end{tabular}
    }
    \vspace{-10pt}
    \label{tab:datasets}
\end{table}

\section{prompt illustration}
\label{app: prompt}
\begin{figure}[t]
\centering
\includegraphics[width=0.48\textwidth]{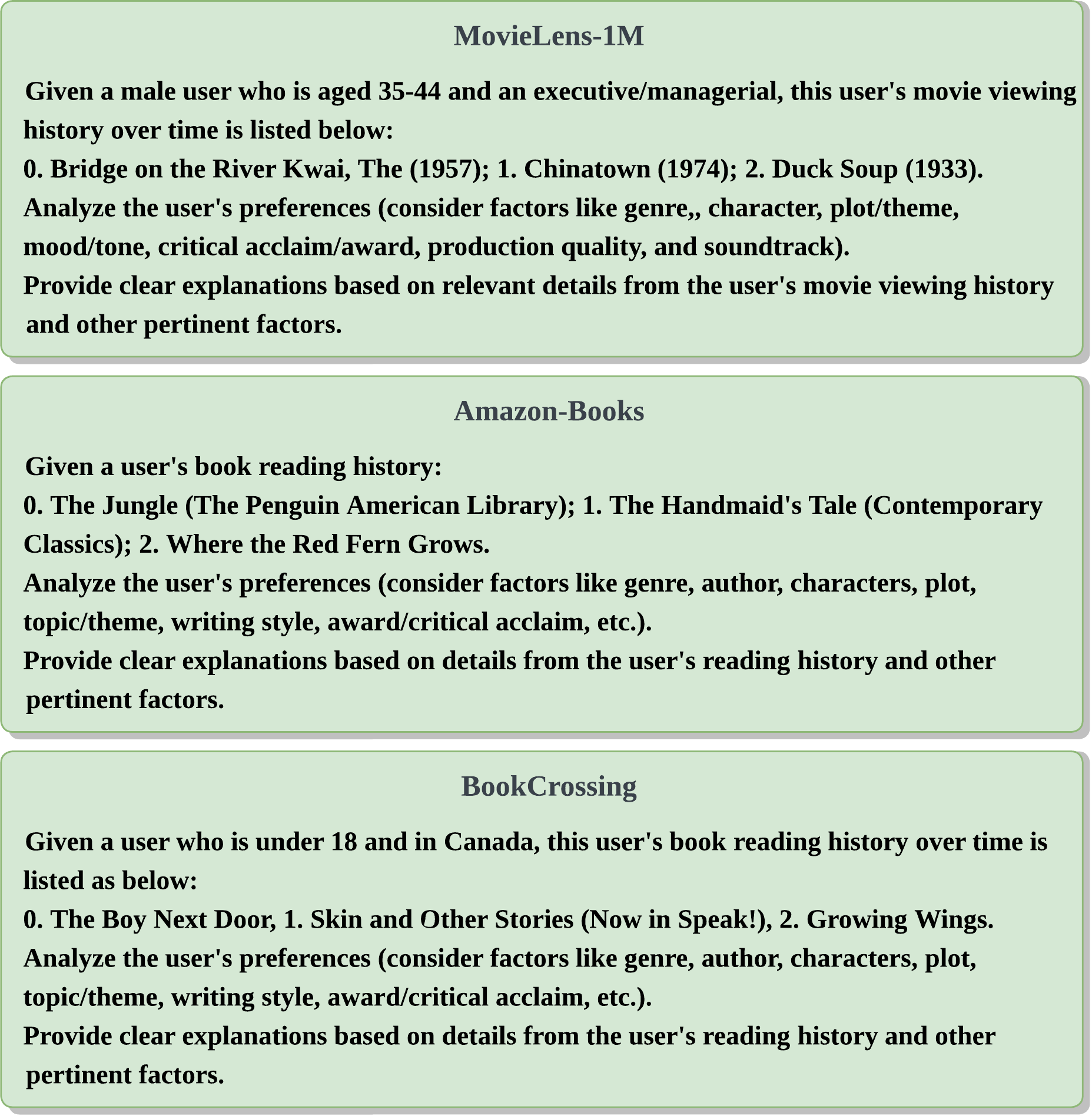}
  \caption{Examples of user prompt templates on three datasets for \textit{Semantic Vector Generation}.
  }
  \label{fig: user prompt}
\end{figure}

\begin{figure}[t]
\centering
\includegraphics[width=0.48\textwidth]{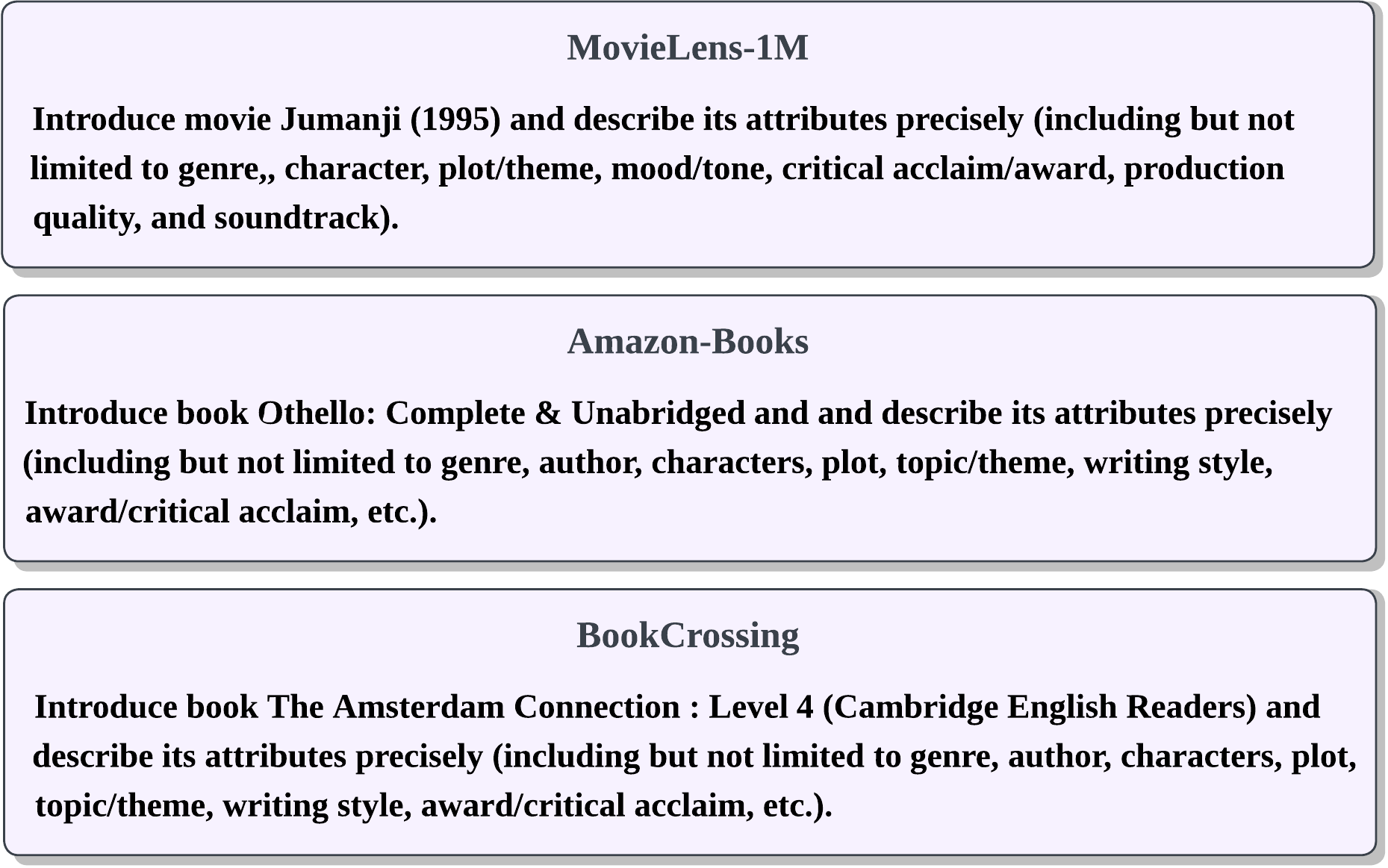}
  \caption{Examples of item prompt templates on three datasets for \textit{Semantic Vector Generation}.
  }
  \label{fig: item prompt}
\end{figure}

We demonstrate several examples to illustrate the user and item prompt templates used for \textit{Semantic Vector Generation} in Section~\ref{sec:Semantic Vector Generation} on the three datasets. 
Figure~\ref{fig: user prompt} shows the examples of user prompt templates on three datasets, which is composed of the user profile, the earliest interaction sequence and possible analyzing aspects. Similarly, Figure~\ref{fig: item prompt} illustrates the item prompt examples for knowledge generation in \textit{Semantic Vector Generation}.

\section{Implementation details}
\label{app: implementation}
As a model-agnostic framework, we jointly train the GNN weights in our AutoGraph with the downstream backbone recommendation models. The loss function is the common cross-entropy loss for sequential recommendation~\cite{boka2024survey}.  AdamW~\cite{loshchilov2017decoupled} is adopted as the optimizer. We optimize the backbone models with the learning rate chosen from \{1e-3, 2e-3, 1e-2, 2e-2\}, the training batch size from \{64, 128, 256\} and  weight decay from \{1e-3, 1e-2\}. The training processes have a maximum epoch of 100 with a early stop patience of 3 epochs. Other model-specific hyperparameters (\eg , the number of GRU layers, the number of attention heads) are also determined
through grid search to achieve the best results. All the experiments are conducted on V100 GPUs.

Moreover, we choose the open-source Vicuna-7B-v1.5~\cite{vicuna2023} as the LLM for semantic knowledge acquiring, and the representation for the generated text is obtained by averaging pooling over the last hidden layer of the LLM. 
For fair comparison, the \textit{LLM-augmented} baselines (\ie , KAR and UIST) share the same generated textual knowledge with \modelname. Besides, UIST uses the same residual quantization configuration as our framework. 

Next we describe the details of quantization process in our framework, and hyperparameter configurations for different backbone models and baseline methods.

\subsection{Residual Quantization}
\label{app: RQ}
For residual quantization configurations, the number of codebooks is set to 3 and the dimension of codebook vector is set to 32. Both the encoder and decoder are multi-layer perceptrons (MLPs) with \{512, 256\} as hidden dimensions. For user-side and item-side quantization, each codebook has 300 and 4096 vectors on BookCrossing dataset, 300 and 256 vectors on MovieLens-1M dataset, and 300 and 512 vectors on Amazon-Books dataset respectively. 
Moreover, to prevent the codebooks from collapse, in which case only a few codebook vectors are used, we follow previous works~\cite{rajput2024recommender, hou2023learning} and apply K-means clustering on the first training batch, then the codebooks are initialized with the clustering centroid vectors. 
Other training hyperparameters are chosen to obtain a high codebook active ratio above 90\%.

\subsection{Backbone Models}

\begin{itemize}[leftmargin=10pt]
    \item \textbf{YouTubeDNN}~\cite{covington2016deep}. On the three datasets, the DNNs are MLPs with the ReLU activation function. The number of layers is chosen from \{2, 3\}. The hidden layer dimensions of MLPs are halved at each layer, with the final output dimension mapped to 32.
    \item \textbf{MIND}~\cite{li2019multi}. The number of interest extracted by the capsule network is set to 3. The iteration time of the \textit{Dynamic Routing} proposed in~\cite{li2019multi} is set to 3.
    \item \textbf{GRU4Rec}~\cite{hidasi2015session}. The number of GRU layers is selected from \{1, 2, 3\}. The dimension of the hidden state is set to 32.
    \item \textbf{SASRec}~\cite{kang2018self}. The number of attention layers is chosen from \{1, 2, 3\}. The number of attention heads is selected from \{1, 2, 4\}. The attention hidden size is set to 32.
\end{itemize}

\subsection{Baseline Methods}
The baseline methods can be divided into two categories: (1) \textit{LLM-augmented methods} and (2) \textit{graph-enhanced methods}. 

\begin{itemize}[leftmargin=10pt]
    \item \textit{LLM-augmented methods}. For fair comparison, \textbf{KAR} and \textbf{UIST} use the same generated semantic vectors as our AutoGraph framework. Moreover, \textbf{UIST} shares the quantization configurations with AutoGraph, as is illustrated in Section~\ref{app: RQ}. We use 2-layer MLPs to incorporate the augmented vectors from \textbf{KAR} and \textbf{UIST} into the backbone models.
    
    \item \textit{Graph-enhanced methods}. For \textbf{LightGCN}, we follow the original paper~\cite{he2020lightgcn} and use GCN as the message aggregator. For \textbf{CCGNN} and \textbf{TopoI2I}, we use GAT as our AutoGraph for fair comparison. The number of graph layers is selected from \{1, 2, 3\}. Neighbor sampling is used, with the node degree selected in a range from 12 to 18. The hidden size of GNNs is set to 32. The number of GAT attention heads is selected from \{1, 2, 4\} for \textbf{CCGNN} and \textbf{TopoI2I}. The NER model chosen for \textbf{CCGNN} is based on BERT-base~\cite{devlin2018bert}. For \textbf{TopoI2I}, we follow the original prompt template used in~\cite{sun2023large} and set the number of in-context demonstrations to 4 for LLMs to infer pairwise item similarity.
\end{itemize}

\section{efficiency evaluation setting}
\label{app: efficiency}
The time is computed on a single V100 GPU. For CCGNN, we choose a NER model based on BERT-base~\cite{devlin2018bert}. For TopoI2I and our \modelname, we experiment on Vicuna-7b-v1.5~\cite{vicuna2023}. We use vLLM~\cite{kwon2023efficient} for LLM inference of a batch size 1. No quantization or other inference techniques are used. For \modelname, the running time in initial graph construction is the sum of time cost of both user graphs and item graphs, while in incremental node insertion the time cost is averaged for user nodes and item nodes. Moreover, the runtime reported in Table~\ref{tab:efficiency} includes the processing time following LLM invocation. However, the overall computational cost is dominated by the LLM usage itself, while the additional processing introduces only negligible overhead and thus does not significantly affect the total runtime.

Next, we provide more detailed analysis on the time complexity of our AutoGraph framework. Under the definition of big $O$ notation, our pointwise method does have the $O(N)$ complexity for initial graph construction.
\begin{itemize}[leftmargin=10pt]
    \item \textbf{Theoretical Analysis.} The overall running time for graph construction is given by $ Nt_1 + m $, where $ N $ is the number of nodes (i.e., LLM calls), $ t_1 $ is the time required for a single LLM call per node, and $ m $ denotes the time for codebook training. Assuming a batch size of $ a $ and training epochs of $ e $, then $ m = \frac{Net_2}{a}$, where $ t_2 $ is the time of one forward and backward pass of a training batch for RQ-VAE. Note that big $O$ notation is concerned with the rate of growth of a function, discarding constant multiplicative factors and lower-order terms~\cite{cormen2022introduction,sipser1996introduction}. Hence, we have $O(Nt_1 + m) = O(N(t_1+\frac{et_2}{a})) = O(N)$.

    \item \textbf{Empirical Analysis.} Even in practice, $ t_1 $ is more than an order of magnitude larger than $ \frac{et_2}{a} $. For instance, on the ML-1M dataset, $ t_1 \approx 18.50\,\text{s} $, whereas with $ e=3 $ and $ a=512 $, $ \frac{et_2}{a} \approx 0.05\,\text{s} \ll t_1 $.
    
\end{itemize}

\section{More experiments}
\label{sec: more experiments}

\subsection{The strategy for generating semantic factors}

\begin{table}[t]
\caption{Ablation study w.r.t. different strategies for generating latent semantic factors. The best result is given in bold, and the second-best value is underlined.
}
\vspace{-10pt}
\label{tab:ID strategy}
\resizebox{\linewidth}{!}{
\renewcommand\arraystretch{1.1}
\begin{tabular}{c|c|cccc|cccc}
\toprule
\hline
 \multicolumn{1}{c|}{\multirow{2}{*}{\makecell{Backbone\\ Model}}} & \multicolumn{1}{c|}{\multirow{2}{*}{Strategy}}  &  \multicolumn{4}{c|}{MovieLens-1M} &  \multicolumn{4}{c}{Amazon-Books} \\ 
 \multicolumn{1}{c|}{} & \multicolumn{1}{c|}{} &  NG@10 & HR@10 & MRR & GAUC &  NG@10 & HR@10 & MRR & GAUC  \\ \hline
 
\multicolumn{1}{c|}{\multirow{3}{*}{YT-DNN}} & LSH & 0.0560 & 0.1045 & 0.0544 & 0.9040 & 0.0486 & 0.0956 & 0.0451 & 0.8195 \\
& HC & \underline{0.0578} & \underline{0.1105} & \underline{0.0557} & \underline{0.9103} & \underline{0.0552} & \underline{0.1077} & \underline{0.0501} & \underline{0.8338} \\
& RQ (Ours) & \textbf{0.0707} & \textbf{0.1221} & \textbf{0.0686} & \textbf{0.9128} & \textbf{0.0609} & \textbf{0.1173} & \textbf{0.0545} & \textbf{0.8362} \\
 \hline

\multicolumn{1}{c|}{\multirow{3}{*}{MIND}} & LSH & \underline{0.0469} & \underline{0.0946} & \underline{0.0461} & \underline{0.9053} & \underline{0.0592} & \underline{0.1153} & 0.0516 & 0.8105       \\
& HC & 0.0440 & 0.0861 & 0.0396 & 0.9021 & 0.0574 & 0.1112 & \underline{0.0518} & \underline{0.8106}  \\
& RQ (Ours) & \textbf{0.0643} & \textbf{0.1166} & \textbf{0.0622} & \textbf{0.9126} & \textbf{0.0737} & \textbf{0.1373} & \textbf{0.0656} & \textbf{0.8297} \\
 \hline

 \bottomrule 
\end{tabular}
}
\vspace{-10pt}
\end{table}

We compare the following alternatives with residual quantization (RQ) to generate different levels of latent factors: (1) hierarchical clustering (HC)~\cite{cohen2019hierarchical}, which applies K-means clustering hierarchically on different levels of clusters. 
(2) locality sensitive hashing (LSH)~\cite{jafari2021survey}, which hashes high-dimensional data points by random projections. For fair comparison, the input semantic vectors are shared by HC, LSH and RQ. We also keep other hyperparameters consistent for the three strategies (\eg, the number of levels and indices in each level). 

The result is reported in Table~\ref{tab:ID strategy}. We can observe that RQ generally outperforms LSH and HC. The reason behind can be that RQ brings more global information to the graph construction process than HC and LSH. Specifically, LSH performs hashes on local regions of data, while HC creates partitions in data, thus disrupting the global structures. Both of them operates in a more localized fashion than RQ. In comparison, RQ utilizes and preserves the global information well. Since it is trained in a self-supervised manner and quantizes the residuals, RQ is able to capture and encode high-level, global features in the data, while refining the detailed aspects in a way that doesn't interfere with the overall structure.

\subsection{Hyperparameter Study}
\label{app: hyper study}

\begin{table}[t]
\caption{The performance of using different codebook levels for graph construction. The total number of codebook levels is 3. $N/A$ means vanilla user-item graph. The best result is given in bold, and the second-best value is underlined.
}
    \vspace{-10pt}
\label{tab:level number}
\resizebox{\linewidth}{!}{
\renewcommand\arraystretch{1.1}
\begin{tabular}{c|c|cccc|cccc}
\toprule
\hline
 \multicolumn{1}{c|}{\multirow{2}{*}{\makecell{Backbone\\ Model}}} & \multicolumn{1}{c|}{\multirow{2}{*}{Levels}}  &  \multicolumn{4}{c|}{MovieLens-1M} &  \multicolumn{4}{c}{Amazon-Books} \\ 
 \multicolumn{1}{c|}{} & \multicolumn{1}{c|}{} &  NG@10 & HR@10 & MRR & GAUC & NG@10 & HR@10 & MRR & GAUC \\ \hline
 
\multicolumn{1}{c|}{\multirow{4}{*}{YT-DNN}} & $N/A$ & 0.0426 & 0.0906 & 0.0405 & 0.8963 & 0.0444 & 0.0845 & 0.0410 & \textbf{0.8417}       \\
& 0 & 0.0572 & 0.1039 & 0.0566 & 0.9080 & \underline{0.0570} & \underline{0.1125} & \underline{0.0511} & 0.8347 \\
& 0, 1 & \textbf{0.0707} & \underline{0.1221} & \textbf{0.0686} & \textbf{0.9128} & \textbf{0.0609} & \textbf{0.1173} & \textbf{0.0545} & \underline{0.8362} \\
& 0, 1, 2  & 0.0678 & \textbf{0.1231} & \underline{0.0656} & \underline{0.9101} & 0.0561 & 0.1082 & 0.0510 & 0.8360 \\
 \hline

\multicolumn{1}{c|}{\multirow{4}{*}{MIND}} & $N/A$ & 0.0435 & 0.0931 & 0.0410 & 0.8936 & 0.0532 & 0.1029 & 0.0487 & 0.8077      \\
& 0 & \underline{0.0584} & \underline{0.1092} & \underline{0.0566} & 0.9078 & \textbf{0.0737} & \textbf{0.1373} & \textbf{0.0656} & \textbf{0.8297}  \\
& 0, 1 & \textbf{0.0643} & \textbf{0.1166} & \textbf{0.0622} & \textbf{0.9126} & 0.0643 & 0.1203 & 0.0580 & 0.8019 \\
& 0, 1, 2  & 0.0572 & 0.1078 & 0.0554 & \underline{0.9094} & \underline{0.0701} & \underline{0.1355} & \underline{0.0615} & \underline{0.8283}  \\
 \hline

 \bottomrule 
\end{tabular}
}
\vspace{-10pt}
\end{table}

\begin{figure}[t]
  \centering
  \includegraphics[width=0.48\textwidth]{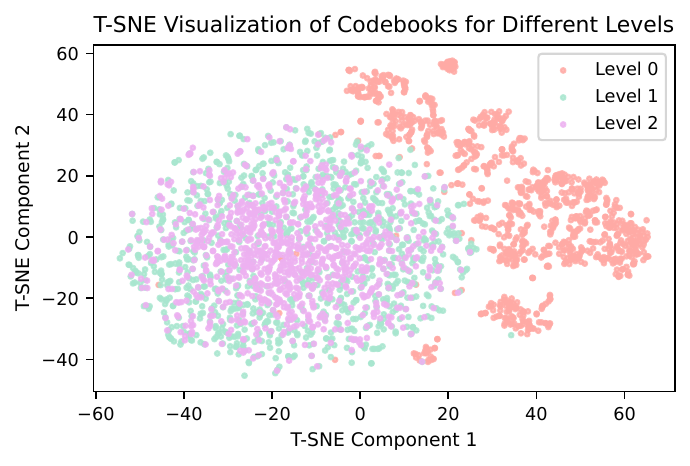}
  \vspace{-15pt}
  \caption{T-SNE~\cite{van2008visualizing} visualization of codebook vectors in different levels on the MovieLens-1M dataset. 
  }
  \vspace{-15pt}
  \label{fig:tsne}
\end{figure}

In this part, we study the number of codebook levels used for graph construction and show the result in Table~\ref{tab:level number}. 
Surprisingly, even though we train 3 levels of codebooks for quantization, including all of them for the final graph construction leads to suboptimal results, while using the first two levels generally yields the best performance. 

For further investigation, we visualize the codebook vectors of different levels in Figure~\ref{fig:tsne}. 
We can observe that codebook vectors of level 0 generally form a compact and informative manifold, while vectors of level 1 and 2 tend to be sparse, uniform and overlapped with each other, which supports the fact that including all three levels leads to suboptimal graph structures. 
The potential reason is that the recursive quantization on vector residuals would encourage codebooks of lower levels to encode more informative and generalized knowledge, and make codebooks of higher levels maintain more noisy and marginal information based on the semantic vectors. 
This highlights that our proposed quantization-based graph construction can filter out the noise in the semantic vectors and improve the overall quality of constructed graph.

\begin{table}[t]
\caption{The performance of using semantic vectors from different text sources. \emph{Original} denotes the simple embeddings of original user/item information. \emph{LLM} denotes the semantic vectors from LLMs. The best result is given in bold.
}
    \vspace{-10pt}
\label{tab: llm importance}

\resizebox{\linewidth}{!}{
\renewcommand\arraystretch{1.1}

\begin{tabular}{c|c|cccc}
\toprule

\hline
Dataset & Text Source & NG@10 & HR@10 & MRR & GAUC \\ \hline
\multirow{2}{*}{MovieLens-1M} & Original & 0.0398 & 0.0876 & 0.0394 & 0.8921 \\
 & LLM & \textbf{0.0707} & \textbf{0.1221} & \textbf{0.0686} & \textbf{0.9128} \\ \hline
\multirow{2}{*}{Amazon-Books} & Original & 0.0438 & 0.0887 & 0.0414 & 0.8001 \\
 & LLM & \textbf{0.0609} & \textbf{0.1173} & \textbf{0.0545} & \textbf{0.8362} \\ \hline
  \bottomrule 

\end{tabular}

}
\vspace{-10pt}

\end{table}

\subsection{The Importance of LLMs in Graph Construction}
We provide experiments to evaluate the information gain introduced by LLMs. Specifically, we use a text encoder (\ie, bge-base-en-v1.5~\cite{bge_embedding}) to embed original item/user descriptions, compared with the LLM-enriched semantic vectors. We leverage YouTubeDNN as the framework, and keep other configurations the same for both types of semantic vectors. The results are shown in Table~\ref{tab: llm importance}, from which we can see that the original semantic information is unilateral and shallow for recommendation. Actually in practice, we find that the simple embeddings from original texts lead to poor quantization quality, and cause codebook collapse. In contrast, the in-depth semantics from LLMs allow for meaningful graph construction, thus improving recommendation.

 \subsection{Case Study}
\label{app: case study}

\begin{figure}[t]
    \vspace{-10pt}
  \centering
  \includegraphics[width=0.45\textwidth]{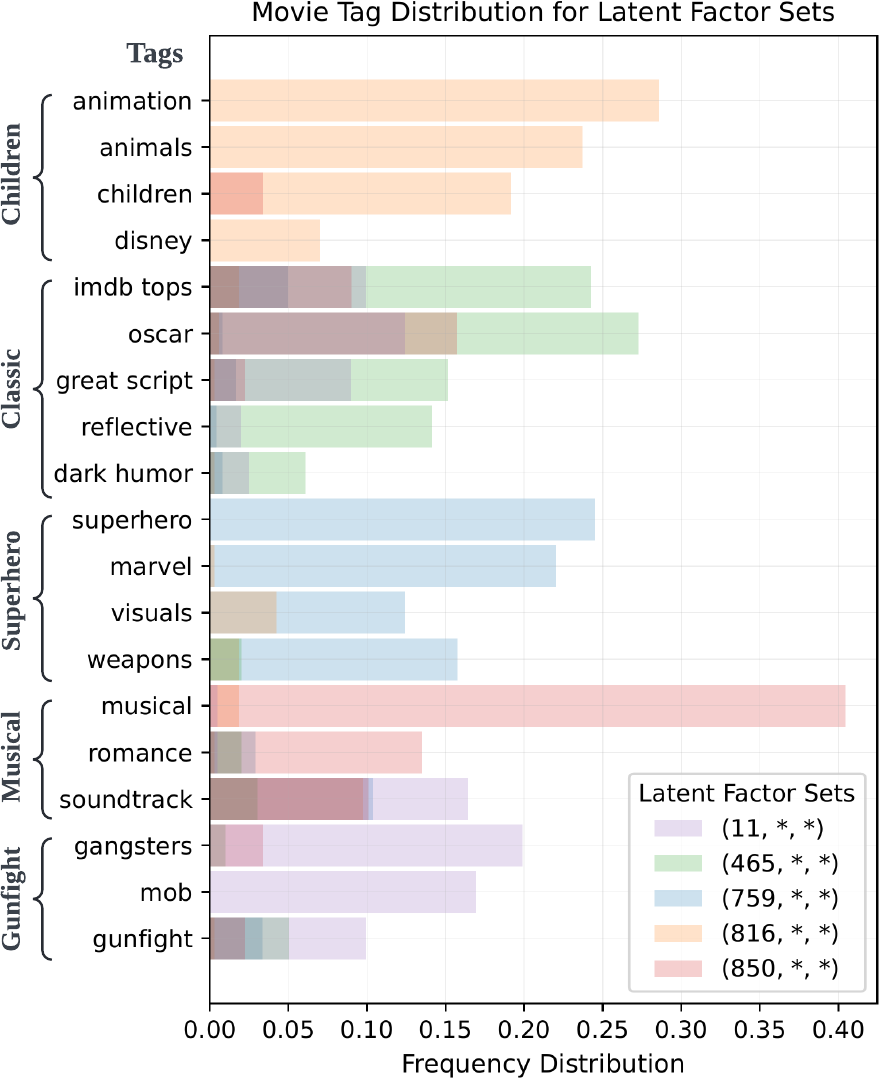}
  \vspace{-10pt}
  \caption{Qualitative case study of the latent semantic factors learned by vector quantization on the MovieLens-1M dataset. We randomly select several latent factor sets and visualize the movie tag distribution of them. The curly bracketed notes on the left are coarse-grained summaries of the fine-grained tags for each factor set.}
  \vspace{-15pt}
  \label{fig:tag distribution}
\end{figure}

We analyze the latent semantic factors learned for the MovieLens-1M dataset. We randomly select several latent semantic factor sets, identify the movies belonging to these factor sets, and collect the tags of the movies online~\footnote{\url{https://movielens.org/}}. 
Then, as shown in Figure~\ref{fig:tag distribution}, we visualize the tag distributions for these latent factor sets.
Each color represents a different factor set sharing the same first-level prefix factor. 
Each bar represents the corresponding tag frequency for a certain factor set. 
The curly bracketed notes on the left are coarse-grained summaries of the fine-grained tags for each factor set. 

From Figure~\ref{fig:tag distribution} we can observe that the fine-grained semantics represented by the latent factor sets are rich and multifaceted, as each set encodes the semantics of multiple tags. Nevertheless, the factor sets still have a prominent theme, since we can clearly infer the specific commonalities of movies within a factor set. For example, factor set (759, *, *) mainly represents \textit{"superhero"}, while at the same time encodes \textit{"marvel"}, \textit{"visuals"} and \textit{"weapons"}. These latent semantic factors explore the underlying structure of the items and provide interpretation for the recommendations to some extent. Taking them as extra nodes, \modelname\ models the item relationships better and enhances the graph topology with the underlying semantic structure, thus improving recommendation performance.

